\newcommand{\tsub}{\textsubscript}
\newcommand{\tsuper}{\textsuperscript}

\newcommand{\sqrts}{$\sqrt{s}$}

\documentclass{PoS}

\usepackage{amsmath}

\title{Recent STAR Jet Results of the High-Energy Spin Physics Program at RHIC}

\ShortTitle{Recent STAR Jet Results of the High-Energy Spin Physics Program at RHIC}

\author{Daniel L. Olvitt Jr. \thanks{for the STAR Collaboration} \\
        Temple University\\
        E-mail: \email{daniel.olvitt@temple.edu}}
        
\abstract{The production of jets from polarized proton+proton collisions at STAR is dominated by quark-gluon and gluon-gluon scattering. The dijet double spin asymmetry (A\tsub{LL}) is sensitive to the polarized gluon distribution ($\Delta g(x)$).  Dijets are also advantageous because the parton momentum fraction, x, of initial partons may be reconstructed to first order from the final state measurements.  Both jet and dijet A\tsub{LL} measurements at $\sqrt{s}$ = 200 GeV have helped to constrain $\Delta g(x)$ in the range $0.05 < x < 0.3$.  In 2012, data were collected at $\sqrt{s}$ = 510 GeV in order to probe lower values of x, these data are consistent with the $\sqrt{s}$ = 200 GeV results in the overlapping $x_T$ region.  Jet and dijet preliminary A\tsub{LL} results have been released and will soon be incorporated into global analyses. In 2013, high luminosity data, with an estimated 250 pb$^{-1}$ of integrated luminosity were collected at $\sqrt{s}$ = 510 GeV.  These data have a figure of merit of $\sim$3 times that of the 2012 data.  An update on the dijet A\tsub{LL} measurement will be presented using polarized p+p data collected at STAR during 2013. }

\FullConference{XXV International Workshop on Deep-Inelastic Scattering and Related Subjects\\
		3-7 April 2017\\
		University of Birmingham, UK}

\begin{document}

\section{Introduction}

It has been known since the late 1980s that the quark's contribution to the spin of the proton is roughly 30\% \cite{Jaffe}.  Since the EMC results showed that quarks aren't the only contributor to the spin of the proton, determining how much the gluon contributes has been an important question in the spin community.  Colliding high energy polarized proton+proton beams at BNL at the Relativisitic Heavy Ion Collider (RHIC) has given new insight into the polarized gluon distribution ($\Delta$g(x)).  The STAR experiment has collected data from polarized proton collisions at center-of-mass energies $\sqrt{s}$ = 200 GeV and $\sqrt{s}$ = 510 GeV in recent years, and these data have made a strong impact on global fits from the DSSV and NNPDF groups.  With the inclusion of the 2009 RHIC proton+proton data at $\sqrt{s}$ = 200 GeV, including the STAR inclusive jet double-spin asymmetry (A\tsub{LL}), the DSSV group found $\int_{0.05}^1 \Delta g(x,Q^2 = 10\text{ GeV} ^2)dx = 0.20^{+0.06}_{-0.07}$ \cite{DSSV14}.  An independent analysis from the NNPDF collaboration found $\int_{0.05}^{0.5} \Delta g(x,Q^2 = 10\text{ GeV}^2)dx= 0.23^{+0.07}_{-0.07}$ \cite{NNPDFpol1.1}.  Both results have cited the need for more data at lower x values.  The 2013 data sample with an integrated luminosity of $\sim$250 pb\tsuper{-1} \cite{RHIC_SPIN} of polarized proton+proton collisions at $\sqrt{s}$ = 510 GeV will help to probe $\Delta g(x)$ at lower values of the parton momentum fraction, x with reduced uncertainties.  



\section{Inclusive Jet Results}


STAR is uniquely suited to probe the polarized ($\Delta g(x)$) and unpolarized gluon distribution ($g(x)$).  At both \sqrts = 200 GeV and \sqrts = 500 GeV, the hard scattering is dominated by quark-gluon and gluon-gluon scatterings.  Combined with full azimuthal coverage for the Barrel Electromagnetic Calorimeter (BEMC) and Time Projection Chamber (TPC) allows the STAR experiment to perform jet measurements sensitive to $g(x)$ and $\Delta g(x)$. Beginning with the 2009 proton+proton data, the STAR collaboration made the switch from the midpoint cone algorithm to the anti-k\tsub{t} algorithm \cite{IncAll_2009}. This change was made, in part because the anti-k\tsub{t} algorithm is less susceptible to underlying event (UE) and pile-up compared to the midpoint cone algorithm \cite{anti-kt}.

\begin{figure}
\centering
\includegraphics[width=0.6\textwidth]{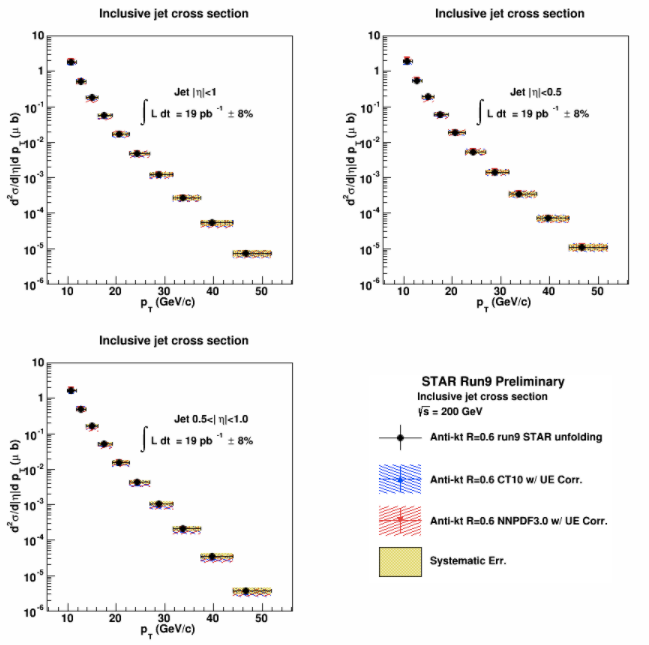}
\caption{Inclusive jet cross section for the 2009 data, with an integrated luminosity of 19 pb\tsuper{-1} at \sqrts = 200 GeV plotted against the particle level jet p\tsub{T}. Top left shows the cross section for $|\eta$ $<$ 1, top right for $|\eta|$ $<$ 0.5, and bottom left for 0.5 $<$ $\eta$ $<$ 1.0. }
\label{fig1}
\end{figure}

One of the goals of the STAR spin program is to better constrain the unpolarized gluon distribution at higher values of x > 0.1.  One observable capable of supplying data in this region is the 2009 inclusive jet cross section \cite{IncXsec_2009} shown in figure \ref{fig1}.  The observable is measured in different $\eta$ bins in order to help contrain the unpolarized gluon distribution for different values of x.  The top left plot of figure \ref{fig1} shows the full mid-rapidity range for $|\eta|$ $<$ 1, while the bottom left is for 0.5 $<$ $|\eta|$ $<$ 1 and the top right plot shows a tighter mid-rapidtiy region of $|\eta|$ $<$ 0.5.  For the 2009 data, an integrated luminosity of 19 pb\tsuper{-1}, with a luminosity uncertainty of 8\% was determined by the beam-beam-counters (BBC).  The main systematic error comes from the calibration of the electromagnetic calorimeter.  Corrections were made for hadronization and UE for each jet p\tsub{T} bin, the total correction is approximately 5\%.  A comparison to Next-to-Leading Order (NLO) perturbative QCD (pQCD)  theory was made using the CT10 \cite{CT10} and the NNPDF3.0 \cite{NNPDF3.0} PDF sets, and the data agree well with both PDF sets.  


\begin{figure}
\centering
\includegraphics[width=0.6\textwidth]{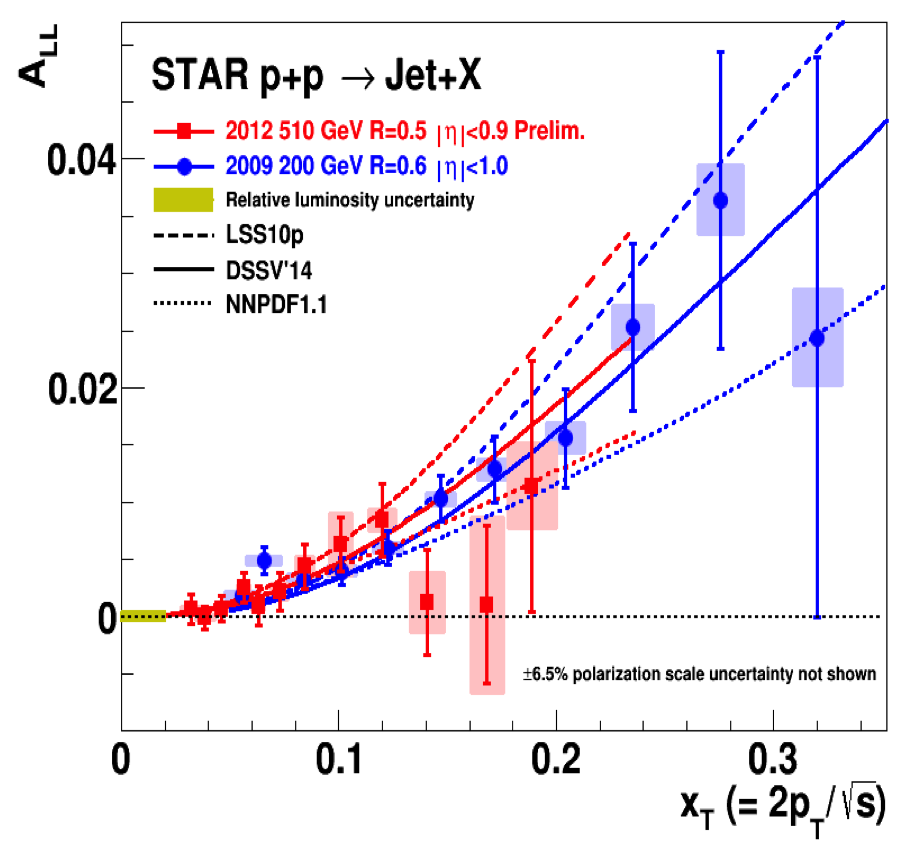}
\caption{Inclusive jet double-spin asymmetry A\tsub{LL} for the 2012 data (red), along with the 2009 data (blue).  Both data sets are compared to the DSSV14 and the NNPDFpol1.1 global fits.  These global fits agree with the data.}
\label{fig2}
\end{figure}

Along with measuring the unpolarized gluon distribution, the STAR experiment is able to probe the polarized gluon distribution.  The 2009 inclusive jet double spin asymmetry (A\tsub{LL}) \cite{IncAll_2009} data were included into both the global fits of the DSSV \cite{DSSV14} and the NNPDF \cite{NNPDFpol1.1} groups.  The latest global fits now show strong evidence of a non-zero gluon polarization in the RHIC range with the STAR detector, and the gluon's contribution to the spin of the proton is of the same order as the contribution made by quarks.  In 2012 STAR collected polarized proton+proton data at \sqrts = 510 GeV. The 2012 A\tsub{LL} \cite{Chang} is shown in figure \ref{fig2}, along with the 2009 inclusive jet A\tsub{LL}, since the data sets are measured at different energies, the asymmetries are plotted against x$_T$.  The measurement at higher center-of-mass energy will allow further constraint to $\Delta g(x)$ at lower x values after inclusion into global fits.

\section{Dijet Results}

Dijet measurements at STAR offer the unique opportunity not only to constrain the value of $\Delta g(x)$, but also the functional form versus x \cite{ColdQCD}.  Recently  dijet cross section and A\tsub{LL} results using the 2009 data at \sqrts = 200 GeV were published \cite{PageDijet} .  These results are the first dijet results that STAR has published.   The cross section data are well described by NLO pQCD theory, just like the 2009 inclusive jet cross section.  The 2009 dijet A\tsub{LL} analysis was separated in two different $\eta$ regions, both of which show agreement with DSSV14 \cite{DSSV14} and NNPDFpol1.1 \cite{NNPDFpol1.1}.  Seperating the data into different $\eta$ ranges allows the gluon polarization to be probed in a  symmetric ($\langle$x\tsub{1}$\rangle$ $=$ $\langle$x\tsub{2}$\rangle$) and asymmetric ($\langle$x\tsub{1}$\rangle$ $>$ $\langle$x\tsub{2}$\rangle$ or $\langle$x\tsub{1}$\rangle$ $<$ $\langle$x\tsub{2}$\rangle$) way which will help to better constrain $\Delta g(x)$ over a wider range in x \cite{Surrow}.

\begin{figure}
\centering
\includegraphics[width=0.6\textwidth]{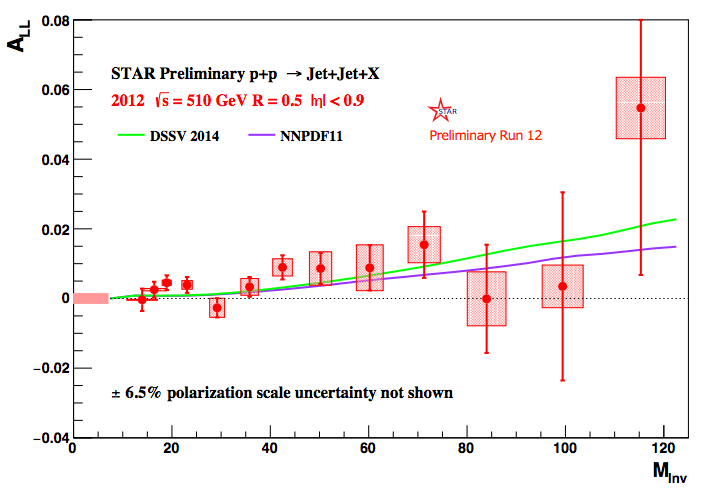}
\caption{Dijet double-spin asymmetry A\tsub{LL} for the 2012 data at \sqrts = 510 GeV, along with a comparison to DSSV14 and NNPDFpol1.1.  Both the data and the global fits are consistent, supporting a positive gluon polarization.}
\label{fig3}
\end{figure}

An A\tsub{LL} analysis was performed for the 2012 data at \sqrts = 510 GeV.  The advantage to measuring the asymmetry for dijets is that the invariant mass is proportional to the combination of the momentum fractions ($M=\sqrt{s} \cdot \sqrt{x_1x_2}$).  At higher center-of-mass energies STAR is able to probe lower x values.  The results of the 2012 dijet A\tsub{LL} \cite{DijetAll_2012} are shown in figure \ref{fig3}.  The dijet A\tsub{LL} was compared to DSSV14 \cite{DSSV14} and NNPDFpol1.1 \cite{NNPDFpol1.1} which was constrained by the run 9 inclusive A\tsub{LL} \cite{IncAll_2009}, and shows good agreement with the global fits.

\begin{figure}
\centering
\includegraphics[width=0.6\textwidth]{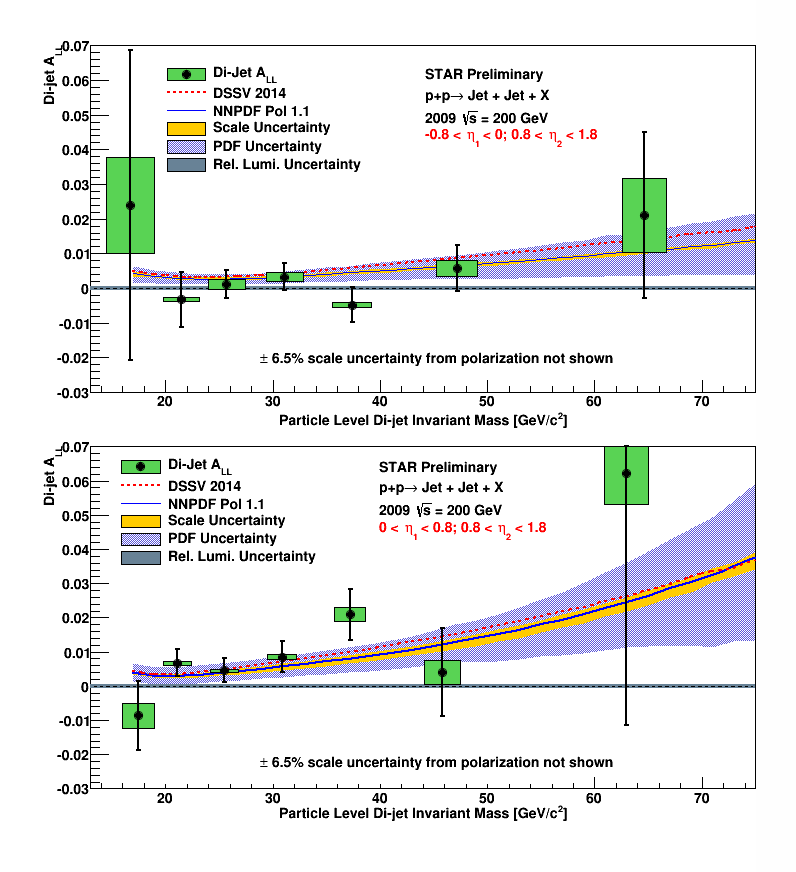}
\caption{Dijet double-spin asymmetry A\tsub{LL} at \sqrts = 200 GeV, compared to the DSSV14 and the NNPDFpol1.1 global fits.  Top plot for -0.8 $<$ $\eta_1$ $<$ 0 and 0.8 $<$ $\eta_2$ $<$ 1.8, bottom plot for 0 $<$ $\eta_1$ $<$ 0.8 and 0.8 $<$ $\eta_2$ $<$ 1.8.  In both $\eta$ ranges the data agree with the global fits. }
\label{fig4}
\end{figure}

\begin{figure}
\centering
\includegraphics[width=0.6\textwidth]{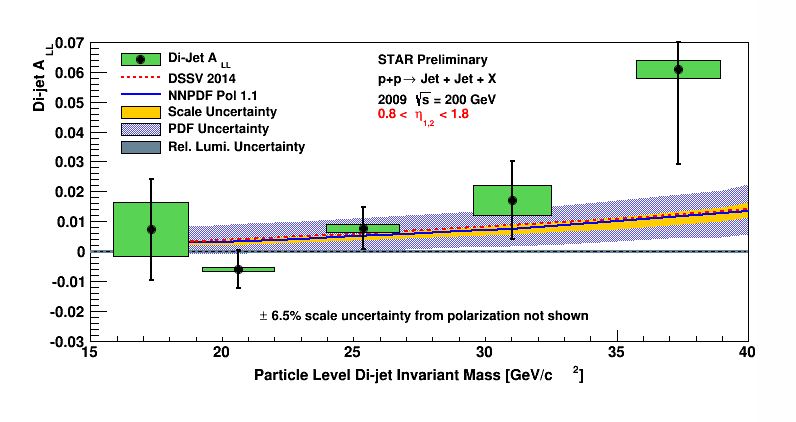}
\caption{Dijet double-spin asymmetry A\tsub{LL} at \sqrts = 200 GeV for 0.8 $<$ $\eta_{1,2}$ $<$ 1.8, as in figure \ref{fig4} the data agree with global fits from DSSV14 and NNPDFpol1.1}
\label{fig5}
\end{figure}

In the last year, STAR has released the first jet analysis that utilized the Endcap Electromagnetic Calorimeter (EEMC).  The 2009 forward dijet A\tsub{LL} \cite{ForwardDijetAll_2009}  utilizes the same data as the published 2009 mid-rapidity analysis \cite{PageDijet}, but is divided into three bins based on $\eta$ where at least one jet of the pair is required to be in the EEMC.  The asymmetry is shown in figures \ref{fig4} and \ref{fig5}.  In figure \ref{fig4} the top plot shows -0.8 $<$ $\eta_1$ $<$ 0 and 0.8 $<$ $\eta_2$ $<$ 1.8, bottom plot shows 0 $<$ $\eta_1$ $<$ 0.8 and 0.8 $<$ $\eta_2$ $<$ 1.8, figure \ref{fig5} shows the $\eta$ range for 0.8 $<$ $\eta_{1,2}$ $<$ 1.8.  In figure \ref{fig5} requiring both jets to be in the EEMC represents the lowest value of x $\sim$0.01 that can be probed at \sqrts = 200 GeV with current STAR detector subsystems.  Similar to all the previous jet and dijet A\tsub{LL} measurements the forward dijet A\tsub{LL} is consistent with the DSSV14 \cite{DSSV14} and the NNPDFpol1.1 \cite{NNPDFpol1.1} global fits, giving more support to a positive value of $\Delta g(x)$ at lower values of x.  


\section{Summary}

Recent jet and dijet results have probed the polarized and unpolarized gluon distribution function with high precision based on measurements from data collected at STAR.  The 2009 inclusive jet and dijet cross section results at \sqrts = 200 GeV were compared to NLO  pQCD theory, the dijet cross section agreed with the CT10 PDF set \cite{CT10}, while the inclusive jet cross section was compared to CT10 and NNPDF3.0 \cite{NNPDF3.0}, and agreed well with both PDF sets.  Inclusion of the 2009 inclusive jet A\tsub{LL} into independent global fits by the DSSV \cite{DSSV14} group and the NNPDF \cite{NNPDFpol1.1} collaboration showed for the first time a statistically significant non-zero value for the gluon polarization, and that the gluon contributes at the same level as the quark towards the total spin of the proton.  Both these global fits have been compared to the 2009 dijet A\tsub{LL} at mid-rapidity and forward rapidity showing agreement between data and global fits.  The 2012 data collected at \sqrts = 510 GeV were analyzed for an inclusive jet and dijet A\tsub{LL} measurement, which like the run 9 results agrees well with both global fits.  In 2013, STAR collected approximately three times the data at \sqrts = 510 GeV compared to 2012 data.  These data will help to better constrain the value of $\Delta g(x)$, and dijet A\tsub{LL} results are essential to help constrain the functional dependence of $\Delta g(x)$.  Extending the results to the EEMC region of STAR will probe the polarized gluon distribution to x value lower than 0.01.  If a forward tracking and calorimeter (2 $<$ $\eta$ $<$ 4) upgrade were installed, STAR dijet measurements  will access x $\sim$0.001, and provide new insight before the construction of an Electron-Ion Collier (EIC) \cite{Surrow_forward}.


\end{document}